\documentclass[preprint,showpacs,preprintnumbers,amsmath,amssymb]{revtex4}
\usepackage{graphicx}% Include figure files
\usepackage{rotating}
\usepackage{epsf}
\usepackage{dcolumn}% Align table columns on decimal point
\begin{document}
\preprint{preprint - vortex group, iitb/iitp}
\title{Growth, Characterization, Vortex Pinning and Vortex Flow Properties of Single Crystals of Iron Chalcogenide Superconductor FeCr$_{0.02}$Se}
\author{Anil K. Yadav$^1$, Ajay D. Thakur$^2$ \footnote{Corresponding Author \\ Email: ajay@iitp.ac.in}, C. V. Tomy$^1$ \footnote{Corresponding Author \\ Email: tomy@phy.iitb.ac.in}}
\affiliation{$^1$ Department of Physics, Indian Institute of Technology Bombay, Mumbai 400076, India\\
$^2$ School of Basic Sciences, Indian Institute of Technology Patna, Patna 800013, India}
\date{\today}
\begin{abstract}
We report the growth and characterization of single crystals of iron chalcogenide superconductor FeCr$_{0.02}$Se.  There is an enhancement of the superconducting transition temperature (T$_{\rm c}$) as compared to the T$_{\rm c}$ of the single crystals of the parent compound Fe$_{1+x}$Se by about 25\%.  The superconducting parameters such as the critical fields, coherence length, penetration depth and the Ginzburg-Landau parameter have been estimated for these single crystals.  Analysis of the critical current data suggests a fluctuation in electronic mean free path induced ($\delta l$) pinning mechanism in this material. Thermally activated transport across the superconducting transition in the presence of external magnetic fields suggests a crossover from a single vortex pinning regime at low fields to a collective flux creep regime at higher magnetic fields. The nature of charge carriers in the normal state estimated from the Hall effect and thermal transport measurements could provide crucial information on the mechanism of superconductivity in Fe-based materials.
\end{abstract}
\maketitle

Iron chalcogenide $\alpha$-FeSe crystallizes in the hexagonal NiAs type crystal structure with $P6_3/mmc$ space group, and is non-superconducting. A slight excess of Fe stabilizes the superconducting tetragonal phase $\beta$-Fe$_{1+x}$Se (anti-PbO type; space group $P4/nmm$) with a superconducting transition temperature (T$_{\rm c}$) of $\sim 8.5$\,K \cite{hsu1}. FeSe and its derivatives (commonly known as the `11' system among the Iron based superconductors) have drawn immense attention \cite{fese2, fese3} akin to the remarkable similarity in their Fermi surface with that of the FeAs based iron pnictide superconductors \cite{fermi4}.  Therefore, despite their structural and compositional simplicity compared to the pnictide counterparts, they are conceived as the promising model systems to understand the physics of Fe-based superconductors. Structurally, FeSe comprises of stacked layers of corner sharing FeSe$_4$ tetrahedra similar to the FeAs based materials. However, the spacer layers are absent in the Fe-11 system. Analogous to the members of the pnictide family, $\beta$-Fe$_{1+x}$Se also undergoes a structural transition below 100\,K (tetra--ortho) resulting in the distortion of the FeSe layers \cite{str5}. The properties of the off-stoichiometric $\beta$-Fe$_{1+x}$Se (or, $\beta$-FeSe$_{1-x}$)  was observed to be very sensitive to the ambient pressure. Whereas, Mizuguchi {\it et al.} \cite{mizu5} reported a T$_{\rm c}$ of 27\,K at 1.48\,GPa in $\beta$-Fe$_{1+x}$Se, Medvedev {\it et al.} \cite{medv6} found a non-monotonicity in the  evolution of T$_{\rm c}$ with pressure upto a maximum  of 36\,K at 8.9\,GPa. The T$_{\rm c}$  was then seen to decrease within the pressure range 8.9--32\,GPa. A non-monotonic variation of T$_{\rm c}$  was also observed in the high pressure studies by Margadonna {\it et al.} \cite{marg7} where T$_{\rm c}$  was seen to peak at 37\,K at 7\,GPa. Garbarino {\it et al.} \cite{garb8}, on the other hand, showed a monotonic increase in T$_{\rm c}$ with pressure, where, an orthorhombic high pressure phase was seen to develop at 12\,GPa, the  T$_{\rm c}$  reaching a maximum of 34\,K at 22\,GPa. Although the precise nature of the superconducting phase and the exact role played by external pressure is not very clear, the sensitive dependence of T$_{\rm c}$ on ambient pressure suggests a possibility of increasing the T$_{\rm c}$ via the chemical pressure route. Several groups around the world have tried different substitutions  both at the Se site \cite{subs9, subs10,subs10a} and the Fe site \cite{fese2, subs10a, subs13, subs13a}. Growth of A$_x$Fe$_2$Se$_2$ (A~$=$~K, Rb) was reported with a T$_{\rm c}$ of $\sim$ 30\,K \cite{kfe2se2a, kfe2se2b, kfe2se2c}. But this belongs to the space group $I4/mmm$ with a ThCr$_2$Si$_2$ structure. There have been no reports suggesting an enhancement of T$_{\rm c}$ for substitutions at the excess Fe site in single crystals of $\beta$-Fe$_{1+x}$Se. Upon Ni substitution at the Fe site, the  superconducting volume fraction was seen to enhance,  but there was no concomitant increase in T$_{\rm c}$ \cite{subs10a}. We have recently reported  a route to enhance the T$_{\rm c}$ by the substitution of Cr instead of excess iron in polycrystalline FeCr$_x$Se samples \cite{aky14}. We have now been successful in growing single crystals of FeCr$_{0.02}$Se. In this paper, we report the detailed superconducting and transport properties in these single crystalline samples.

The single crystals used in this work were prepared using the self flux method for crystal growth. Powders of 4N Fe, Se and Cr  were homogenized in an agate mortar in the required stoichiometric composition (FeCr$_{0.02}$Se), sealed in an evacuated quartz tube (10$^{-6}$ mbar) and preheated at 1050\,$^\circ$C for 24\,h. It was cooled down to 800\,$^\circ$C at a rate of 2\,$^\circ$C/h and then the furnace was switched off. The sample thus obtained was annealed at 360\,$^\circ$C for 36\,h followed by quenching in liquid N$_2$.  The X-ray diffraction (XRD) and energy dispersive X-ray analysis (EDXA) were performed to confirm the structure and elemental composition. DC magnetization and ac susceptibility measurements were performed using a SQUID-Vibrating Sample Magnetometer (SVSM) (Quantum Design Inc.~USA). The amplitude  and the frequency for the ac susceptibility measurements were kept at $h_{\rm ac}=3.5$\,Oe and 211\,Hz, respectively. Electrical and thermal transport measurements were performed using the Physical Property Measurement System (PPMS), Quantum Design Inc.~(USA). 

Figure~1(a) shows the powder X-ray diffraction pattern of the powdered single crystals of FeCr$_{0.02}$Se. A two phase Rietveld refinement was performed using the tetragonal phase $\beta$-Fe$_{1+x}$Se (anti-PbO type; space group $P4/nmm$) and the hexagonal phase $\alpha-$FeSe (NiAs type; space group $P6_3/mmc$) as reference structures (using atomic positions from ICPDS databases). The atomic positions used for the $\beta$-Fe$_{1+x}$Se phase were: (i) Fe: 2a ($3/4$, $1/4$, $0$), (ii) Se: 2c ($1/4$, $1/4$, $z_{Se}$), and (iii) Cr: 4f ($3/4$, $1/4$, $z_{Cr}$). In the case of $\alpha-$FeSe, the atomic positions used were: (i) Fe: 2a ($0$, $0$, $0$), (ii) Se: 2c ($1/3$, $2/3$, $1/4$), (iii) Cr: 2a ($0$, $0$, $0$). For $\beta$-Fe$_{1+x}$Se phase, the occupancy of 2a, 2c and 4f sites were 0.998, 1 and 0.022, respectively and the values of $z_{Se}$ and $z_{Cr}$ were found to be 0.2691(7) and 0.34(2), respectively. The results of the Rietveld refinement for powdered single crystal of FeCr$_{0.02}$Se are summarized in table I where a comparison with a similar analysis for $\beta$-Fe$_{1+x}$Se (with $x=0.01$) is also provided. The lattice parameters almost match with that of $\beta$-Fe$_{1+x}$Se, indicating that the small amount of excess Cr at the Fe site does not affect the crystal structure appreciably.  This in turn implies that the small amount of chemical pressure exerted by the Cr atom (whose ionic radius is bigger than that of the Fe ion) may be responsible for the increase in the observed T$_{\rm c}$ in this compound. The corresponding parameters for K$_{0.8}$Fe$_2$Se$_2$ / K$_x$Fe$_{2-y}$Se$_2$ (space group $I4/mmm$) \cite{kfe2se2a, kfe2se2c} have also been collected in Table I, which gives a clear indication that our single crystals of FeCr$_{0.02}$Se (space group $P4/nmm$) do not fall in the same category as K$_{0.8}$Fe$_2$Se$_2$ / K$_x$Fe$_{2-y}$Se$_2$ (space group $I4/mmm$). The single crystals of FeCr$_{0.02}$Se reported in this paper have layered planes held together by weak van der Waals interaction and can therefore be cleaved relatively easily similar to the FeSe$_{1-x}$Te$_x$ \cite{pdas17}, NbSe$_2$ \cite{paltiel18, xiao19, adt20} and Bi-2212 \cite{ooi21, tamegai22} single crystals. The X-ray diffraction pattern obtained for one such cleaved single crystal piece is shown in the main panel of Fig.~1(b). A prominent (101) peak is observed, indicating that the (101)-axis of the single crystal is perpendicular to the cleaved surface. A weak (201) peak is also observed which is very similar to the results reported by Hu {\it et al} \cite{lei34}. This appears to be a common feature related to specific growth details in this class of materials \cite{lei34}. The right inset in Fig.~1(b) shows the optical micrograph of a typical FeCr$_{0.02}$Se single crystal piece. The nominal composition of our single crystals was also verified using energy dispersive X-ray analysis (EDXA) at various locations on the single crystal pieces. It was not possible to obtain good Laue diffraction patterns for our single crystals due to issues related to X-ray absorption by the constituents. Therefore we resorted to selective area electron diffraction (SAED) using a transmission electron microscope for verifying the quality of our FeCr$_{0.02}$Se single crystals. In the left inset of Fig.~1(b), we show the SAED pattern obtained using a tiny piece of FeCr$_{0.02}$Se single crystal. The sharp diffraction spots indicate the high degree of crystalline nature.

Main panel  of Fig.~2(a) shows the temperature variation of the real and imaginary parts of the ac susceptibility data for a single crystal of FeCr$_{0.02}$Se. The temperature at which $\chi^{\prime\prime}(T)$ goes to zero is marked as the onset temperature for the diamagnetic transition, $T^{\rm on;\chi}_c$. In panel (b) of Fig.~2, we present a comparison of the dc magnetization and the electrical resistance measurements across the superconducting transition in the same FeCr$_{0.02}$Se single crystal. The onset ($T^{\rm on;\rho}_c = 14.8$\,K), the mid-point ($T^{\rm mid;\rho}_c = 12.2$\,K)  and the offset ($T^{\rm off;\rho}_c = 11.1$\,K) temperatures obtained at 90\%,  50\% and 10\%, respectively of the normal state resistivity ($\rho_N$) measured in the absence of an external magnetic field ($H=0$) are marked in Fig.~2(b). The zero resistance temperature ($T^{0;\rho}_{c} = 10.5$\,K) is also marked along with the onset temperature ($T^{\rm on;M}_{c}$) obtained using the dc magnetization measurements. The temperatures for the onset of diamagnetism ($T^{\rm on;\chi}_{c}$) and the zero resistance ($T^{0;\rho}_{c}$) coincide with each other (hallmark of superconductivity).  This temperature is identified as the superconducting transition temperature ($T_{c}$) and has a value of 10.5\,K for our crystals of FeCr$_{0.02}$Se.  

Any practical application of a typical type II superconductor demands an understanding of its vortex phase diagram \cite{wire23}. In particular, it is of utmost importance to obtain the lower critical field, $H_{c1}$ and the upper critical field, $H_{c2}$ values at various temperatures below $T_{c}$. In panels (a) and (b) of Fig.~3 we present the $H_{c2}$ data obtained from the electrical transport measurements via the magnetic field dependence of the $T_{c}^{\rm on;\rho}$, $T_{c}^{\rm mid;\rho}$ and $T_{c}^{\rm off;\rho}$ values (see Fig.~2(b)) for both $H\parallel (101)$ and $H\perp (101)$. The values of the slope ($\frac{dH_{c2}}{dT}$), extracted from the graph are $-$23\,kOe/K, $-$26\,kOe/K and $-$25\,kOe/K, for the values corresponding to $T_{c}^{\rm on;\rho}$, $T_{c}^{\rm mid;\rho}$ and $T_{c}^{\rm off;\rho}$ (identical within the first decimal place for both $H\parallel (101)$ and $H\perp (101)$), respectively. The upper critical field values at zero temperatures  $H_{c2}(0)$ are therefore estimated to be 240~kOe, 220~kOe and 175~kOe (for both $H\parallel (101)$ and $H\perp (101)$) using the Werthamer-Helfand-Hohenberg (WHH) formalism \cite{whh24}, where, $H_{c2}(0)=-0.693\left(\frac{dH_{c2}}{dT}\right)T_{c}$ for the three different criteria, respectively. This may indicate the absence of any anisotropy in the superconducting properties of  FeCr$_{0.02}$Se. The estimated values for $H_{c2} (0)$ are  close to the Pauli paramagnetic limit $H_{P}(0) = 1.84 T_{c}$ \cite{ppl25}, suggesting spin-paramagnetic effect as the dominant pair breaking mechanism similar to the claims made for the FeSe, Fe(Se,Te) and Fe(Te,S) systems  \cite{lei35a, lei35b, lei35}. An estimate of the superconducting coherence length $\xi(0)$ was made using the Ginzburg-Landau expression $H_{c2}(0) = \Phi_{0} / 2\pi \xi^2(0)$. The values for $\xi(0)$ thus obtained are 1.68\,nm, 3.87\,nm and 4.33\,nm from the plots obtained for $T_{c}^{\rm on;\rho}$, $T_{c}^{\rm mid;\rho}$ and $T_{c}^{\rm off;\rho}$, respectively. In order to make an estimate of the lower critical field, $H_{c1}$, the virgin curves (corresponding to the isothermal $M$--$H$ measurements) were obtained after zero field cooling. The criteria of deviation from  linearity of the  virgin $M$--$H$ plots was used to determine the values of $H_{c1}$. In Fig.~4, we plot the $H_{c1}$ data for both $H\parallel (101)$ and $H\perp (101)$. A BCS fit \cite{bcs25a} yields  $H_{c1} (0)$ values of 67\,Oe and 33\,Oe for $H\parallel (101)$ and $H\perp (101)$, respectively. The values of penetration depth $\lambda (0)$, Ginzburg-Landau parameter $\kappa (0)$ along with the critical fields are tabulated in Table II for FeCr$_{0.02}$Se. A comparison is also made in the table with the superconducting parameters of the related Fe11-based superconductors \cite{lei34, kim34, hu34}. It should be noted here that the BCS fit is not very good at temperatures close to $T_{c}(0)$ where the experimental data shows a curvature reminiscent of multi-band superconductors.

In the case of type-II superconductors, it is feasible to obtain a practical estimate of the critical current density ($J_{c}$) via suitable magnetization measurements \cite{bean26, bean27}. This is useful when the contact resistance in a typical magneto-transport measurement is such that passing large currents through the sample for the estimation of $J_{c}$ is impractical without resorting to special protocols \cite{xiao19} for making contacts on the sample. In our case we resort to the contactless technique via magnetization measurements for the determination of $J_{c}(H,T)$ using the Bean's critical state model \cite{bean26, bean27}. Typical  isothermal magnetization hysteresis ($M$--$H$) loops recorded at several temperatures between 2\,K and 8\,K (data at other temperatures not shown here for clarity) are shown in Fig.~5(a) (for $H\parallel (101)$) and Fig.~5(b) (for $H\perp (101)$). The peak in magnetization located at a field value little above the nominal zero field in a given $M$--$H$ loop represents the characteristic first magnetization peak, which amounts to (near) full penetration of the applied field in the bulk of the sample after zero field-cooling. We did not observe any prominent features corresponding to the fishtail effect/second magnetization peak in our crystals in contrast to the observation of such effects in the case of Fe(Se,Te) crystals \cite{pdas17}. The identical nature of the $M$--$H$ curves for both the orientations of the field corroborates the absence of any  anisotropy (see Fig.~3, where, the values of $H_{c2}$ for both $H\parallel (101)$ and $H\perp (101)$ are presented). Making use of the Bean's critical state model formalism \cite{bean26, bean27}, we extracted the critical current density values from the isothermal $M$--$H$ data. Within the Bean's model,  $J_c = 20 \frac{\Delta M}{a(1- \frac{a}{3b})}$ \cite{bean27}, with $a =3.93$\,mm,  $b = 3.47$\,mm being the sample dimensions perpendicular to the field direction and $\Delta M$ is the difference between the magnetization measured during the return and the forward legs of the $M$--$H$ loops. Figure~6 shows the plot of the critical current density $J_{c}(H)$ at temperatures between 2\,K and 10\,K for both $H \parallel (101)$ (panel (a)) and $H\perp (101)$ (panel (b)). The absence of any anomalous features is evident from the plots.

Vortex pinning in type-II superconductors can be broadly classified into two categories, viz., $\delta T_{c}$ pinning (arising because of the spatial fluctuations in the transition temperature, $T_{c}$ across the sample) and $\delta l$ pinning (caused by the spatial variations in the charge carrier mean free path, $l$). Whereas $\delta l$ pinning might arise due to randomly distributed weak pinning centers, the $\delta T_{c}$ pinning arises due to correlated disorder, either naturally occurring in certain materials, or, resulting from heavy ion/proton irradiation \cite{blatter28, prl199429}. In the case of $\delta T_{c}$ pinning, the normalized critical current density, $J_c(t)/J_c(0) = (1-t^2)^{7/6}(1+t^2)^{5/6}$, while for $\delta l$ pinning, $J_c(t)/J_c(0) = (1-t^2)^{5/2}(1+t^2)^{-1/2}$, where $t=T/T_{c}(0)$ \cite{blatter28, prl199429} is the reduced temperature. Following a recipe similar to Das {\it et al} \cite{pdas17},  we make a comparison of the experimentally obtained $J_{c}$ values (for $H \parallel (101)$) with the theoretically expected variations within the $\delta T_{c}$ and $\delta l$ pinning scenarios in Fig.~7. The observations point to the dominance of the $\delta l$ pinning mechanism in FeCr$_{0.02}$Se, suggesting the occurrence of single vortex pinning by randomly distributed weak pinning centers. 

In the presence of an external magnetic field, the dissipation behavior of a typical type-II superconductor is guided by the competition between the pinning forces (due to $\delta l$ or $\delta T_{c}$ pinning) and the Lorentz force (acting on the flux lines because of the transport current). The dissipation behavior comprises of two distinct regimes: (a) the pinning force dominated flux-creep regime, and (b) the Lorentz force dominated flux-flow regime. For the high $T_{c}$ superconductor Bi$_{2.2}$Sr$_2$Ca$_{0.8}$Cu$_2$O$_8$, Palstra {\it et al} \cite{palstra32, palstra33} have formulated a dissipation behavior guided uniquely by an orientation and field dependent activation energy $U_{0} (H,\phi)$. It is of interest to evaluate the extent to which such formalism is applicable for the Fe-based superconductors. The main panel in Fig.~8 shows the variation of resistivity ($\rho$) as a function of temperature for a suitably cleaved single crystal of FeCr$_{0.02}$Se at 0\,kOe and 90\,kOe ($H\parallel$ (101)). The overall nature of the resistivity curves resembles that of the reported behaviour for FeSe compounds; the S-type resistivity variation pointing towards the psuedo-gap nature of the charge carriers \cite{song}. We also notice an anomalous behaviour around 70\,K between the two resistivity curves  (0\,kOe and 90\,kOe); there is a crossover of resistance between the two curves at $\sim 72$\,K (see a portion of the magnetoresistance curve in the inset (a) of Fig.~8). The resistance is higher for 90\,kOe below this temperature, whereas the resistance for 0\,kOe is higher above this temperature. This cross over may be associated with the structural transition \cite{str5} in this compound.  In Fig.~8, the inset panel (b) shows the variation of $\frac{d\rho}{dT}$ with temperature, clearly showing a discontinuity associated with the structural transition. A large magnetoresistance below 72\,K leads to an increased slope of $\frac{d\rho}{dT}$ for 90\,kOe.

In order to obtain the activation energy $U_0$, in Fig.~9, we plot $\ln\rho$ versus $1/T$ at various fields $H$ for both $H\parallel (101)$ (panel (a)) and $H\perp (101)$ (panel (b)). Linear fits are performed on the low temperature part (large $1/T$ behavior) and these are shown by solid lines in both the panels of Fig.~9. The slopes obtained from these fits yield the values of the activation energies $U_0$ at various field values for both $H\parallel(101)$ and $H\perp(101)$. As we know, for thermally activated flux-flow behavior, $\ln\rho (T,H) = \ln \rho_0(H) - U_0(H)/T$ \cite{palstra32}. The plots of $\ln \rho$ versus $1/T$ at various fields have a common intersection point (see panels (a) and (b) of Fig.~9) which should ideally correspond to the $T_{c}$ of the superconducting specimen. In our crystal, this intersection point corresponds  to a temperature of 12.8\,K  ( $1/T \approx 0.078$\,K$^{-1}$) which is indeed very close to the $T_{c}^{\rm mid}$ value obtained from the resistance measurements. The variation of $U_0$ with $H$ (for both $H\parallel (101)$ and $H\perp(101)$) is shown in Fig.~10. We find a very similar behavior for both the field orientations. Power law fits ($\sim H^{-\alpha}$) are performed on the $U_0(H)$ data and the two regimes are clearly visible with a crossover occurring at a characteristic field $H_{\rm cr}$ ($\approx 38$\,kOe) for both $H\parallel (101)$ and $H\perp (101)$. The values for the exponent $\alpha$ in the two field regions for different field orientations are provided in Table III. It should be noted here that a similar analysis for Bi$_{2.2}$Sr$_2$Ca$_{0.8}$Cu$_2$O$_8$ by Palstra {\it et al} \cite{palstra32, palstra33} also yielded a similar crossover field behavior with different values of $\alpha$ on the two sides. However, in their case, very different values for $\alpha$ were obtained for the two field orientations (possibly related to the large anisotropy present in the single crystals of Bi$_{2.2}$Sr$_2$Ca$_{0.8}$Cu$_2$O$_8$). Lei {\it et al} \cite{lei35} have found a similar behavior for single crystals of $\beta$-FeSe with $H_{\rm cr} \approx 30$\,kOe (smaller than the crossover field of $\approx 38$\,kOe observed in our single crystals). A similar crossover behaviour was also reported by Lee {\it et al} \cite{lee35} in the iron arsenide superconductor SmFeAsO$_{0.85}$.  For our FeCr$_{0.02}$Se single crystal, we have obtained  $\alpha \approx 0.37$ for $H<H_{\rm cr}$ while for $H>H_{\rm cr}$, $\alpha \approx 0.88$ ($H\parallel (101)$). The values of $\alpha$ for the other field orientation ($H\perp (101)$) are within a few percent of the values for $H\parallel (101)$. These values for $\alpha$ are slightly larger than the values obtained for $\beta$-FeSe \cite{lei35}. A change in the value of the exponent $\alpha$ by about a factor of 3 across $H_{\rm cr}$ with a relatively small value at low fields suggests a crossover from a single-vortex pinning dominated regime to a collective flux creep regime across $H_{\rm cr}$ \cite{yesh36}.   

The nature of charge carriers in the normal state just prior to the superconducting transition is one of the crucial ingredients that is required to physically understand the mechanism of superconductivity in any material system. In order to develop an understanding in this direction, we have performed both the Hall effect studies and the thermal transport measurements (including the determination of Seebeck coefficient $S$) as a function of temperature in a single crystal piece of FeCr$_{0.02}$Se.  Contacts for the Hall measurements were made in the five probe geometry using silver epoxy and 30 micron gold wires. The observed Hall coefficient ($H = 50$\,kOe) as a function of temperature is shown in panel (a) of Fig.~11. From the data it is evident that the Hall coefficient is negative below 220\,K pointing to a prominence of electron like charge carriers in this region. Figure~11(b) shows a plot of $S(T)$ in applied magnetic field values of  0\,kOe and 50\,kOe. Below the superconducting transition temperature, $S(T)$ is zero as the charge carriers are involved in the formation of  Cooper pairs. The inset panel in Fig. 11 (b) shows the enlarged view of the $S(T)$ data between 5\,K to 18\,K. It should be noted that the $S(T)$ measurements confirm the  $T_{c}$ of the sample to be ~10.5\,K as obtained from the magnetization and electrical transport studies (c.f., Fig.~2). In the presence of an external field of 50\,kOe, the shifting of $T_{c}$ to the lower temperatures is also evident from the $S(T)$ data. Above $T_{c}$, $S(T)$ has a positive value and it crosses  zero at $\sim$ 65\,K. This change of sign is not seen in the  Hall measurements. Above 65\,K the $S(T)$ changes sign and becomes negative and attains a maximum  magnitude of $\sim 26$\,$\mu$V/K at 109\,K. The $S(T)$ data above 109\,K suggests a change in the type of majority charge carriers. The change in the type of charge carrier at this temperature is also observed in the Hall measurement. Thus, from the $S(T)$ data, it is clearly evident that this material has both types of charge carriers. A modulation in the fraction of the two types of charge carriers as a function of temperature is responsible for the non-monotonic variations in $S(T)$. This could be one of the crucial information in theoretically understanding the nature of superconducting state in this class of materials. In Fig.~11(c) we plot the variation of the thermal conductivity $\kappa$ with temperature for the same crystal in 0\,kOe and 50\,kOe. The local maxima and minima in $\kappa(T)$ roughly correspond to the two temperatures where, $S(T)$ displays sign changes. This indicates a plausible connection between thermal conductivity and the nature of charge carriers.

It can be noted that the occurrence of a sign change in the $S(T)$ data at a temperature of $\sim 65$\,K does not seem to reflect in the Hall data. In order to look at this behaviour in more detail, we measured the Hall voltage $V_H$ as a function of externally applied magnetic fields at various temperatures in the normal state. The results of such  measurements are plotted in Fig.~12. There is a clear change in the slope (at the high field end) of  $V_H$ versus $H$ data at $\sim 100$\,K corroborating the observations in Fig.~11(a). In addition to this, in Fig.~12, one can observe a clear non-monotonic variation of $V_H$ as a function of field at temperatures below $\sim 70$\,K with the $V_H$ value attaining a maxima at intermediate fields. At 70\,K, this non-monotonic behaviour disappears. This can therefore rationalize the two sign changes observed in the $S(T)$ data.

The existence of strong correlations accompanied by a non-Fermi liquid behavior has been suggested for iron chalcogenide superconductors based on a variety of theoretical studies \cite{aichhom,craco}.  It has been proposed that the enhancement in the number of thermally excited carriers above the pseudogap temperature results in a concomitant decrease in the magnitude of $S(T)$ \cite{nishi1, lue, nishi2}. The observed decrease in the value of $S(T)$ at  temperatures beyond $\sim 100$\,K (see Fig.~11(b) for details) hints to a possible connection with the pseudogap behavior in our single crystals of FeCr$_{0.02}$Se. This is further corroborated by the presence of an anomalous feature in $\frac{d\rho}{dT}$  (see top inset in Fig. 8) as propounded earlier for cuprates \cite{ando1, ando2}. A very similar observation was made by Song {\it et al} \cite{song} for single crystals of FeSe. 

To summarize, we have grown single crystals of tetragonal phase of FeCr$_{0.02}$Se (anti-PbO type; space group $P4/nmm$) where, 2\% Cr is substituted in excess at the Fe site. These single crystals have a $T_{c}$ which is 25\,$\%$ larger than the parent $\beta$-FeSe. The basic superconducting properties are corroborated using magnetic susceptibility (both ac and dc) as well as electrical transport measurements. An estimate of $J_{c}$ is obtained using isothermal $M$--$H$ data which is also utilized to establish the nature of pinning mechanism in these single crystals. A one-to-one comparison between the experimentally obtained temperature variation of $J_{c}$ with  the corresponding theoretical estimates for $\delta l$ pinning and $\delta T_{c}$ pinning is performed. Based on this, the pinning properties within these single crystals are attributed to the $\delta l$ pinning mechanism which corresponds to the occurrence of single vortex pinning by randomly distributed weak pinning centers. From low field magnetization measurements and magneto-transport studies, fundamental parameters like $H_{c1}$, $H_{c2}$, $\kappa$, $\xi(0)$ and $\lambda(0)$ are obtained which are fundamental to the understanding of magnetic properties of these single crystals. An attempt is made to understand the flux-flow behavior in these single crystals within the thermally activated flux-flow model. The results point to a crossover from a single-vortex pinning dominated regime to a collective flux creep regime across a characteristic field $H_{\rm cr}$ \cite{yesh36} which is practically independent of the magnetic field  direction. Finally, the issue of nature of charge carriers in the normal state of these single crystals is addressed. For this, we resort to Hall effect, thermopower and thermal conductivity measurements which together suggest a modulation in the fraction of the two types of charge carriers with temperature within the normal state.  This could prove to be a vital input for theoretically understanding the nature of the normal state within these classes of superconductors and possibly the physics of the mechanism of superconductivity in these systems.

CVT would like to acknowledge the Department of Science and Technology for partial support through the project IR/S2/PU-10/2006. AKY would like to thank CSIR, India for SRF grant. ADT acknowledges the Indian Institute of Technology, Bombay for partial financial support during part of this work and the Indian Institute of Technology, Patna for seed grant.

\newpage

\begin{table}
%\begin{sidewaystable}
\centering
\caption[Comparison of lattice parameters of the single crystal of FeCr$_{0.02}$Se with other related compounds]{Comparison of lattice parameters of the single crystal of FeCr$_{0.02}$Se with other related compounds.}
\begin{tabular}{| c | c | c | c | c |}
\hline
\hline
Sample & FeCr$_{0.02}$Se & Fe$_{1.01}$Se \cite{lei34, lei35} & Fe$_{1.01}$Se \cite{mcqueen34} & K$_{0.8}$Fe$_2$Se$_2$ / K$_x$Fe$_{2-y}$Se$_2$ \cite{kfe2se2a, kfe2se2c}\\
\hline
\hline
Nature & single crystal & single crystal & polycrystal & polycrystal\\
\hline
Data & Powder XRD & Powder XRD & Neutron diff. & Powder XRD\\
\hline
Crystal Structure & Tetragonal & Tetragonal & Tetragonal & Tetragonal\\
\hline
Type & anti-PbO & anti-PbO & anti-PbO & ThCr$_2$Si$_2$\\
\hline
Space group & $P4/nmm$ & $P4/nmm$ & $P4/nmm$ & $I4/mmm$\\
\hline
$a$ (\AA) & 3.7730 & 3.7622 & 3.7734 & 3.9136\\
\hline
$c$ (\AA) & 5.5241 & 5.5018 & 5.5258 & 14.0367\\
\hline
$c/a$ & 1.4641 & 1.4623 & 1.4644 & 3.5866\\
\hline
$V$ (\AA$^3$) & 78.639 & - & 78.402 & 214.991\\
\hline
Phase (W$\%$) & 84.6 & 90.7 & - & -\\
\hline
Rexp $\%$ & 2.034 & - & - & 2.22\\
\hline
Rpro $\%$ & 2.909 & - & - & 3.26\\
\hline
Rwp $\%$ & 3.760 & - & 6.56 & 5.15\\
\hline
$\chi^2$ & 3.416 & - & 2.117 & 5.38\\
\hline
\hline

%\label{tab:table1}
\end{tabular}
\end{table}
%\end{sidewaystable}

\newpage
\begin{table}[!t]
\begin{center}
\caption[Comparison of superconducting parameters of the single crystal of FeCr$_{0.02}$Se with other related compounds.]{Comparison of superconducting parameters of the single crystal of FeCr$_{0.02}$Se with other related compounds.}
\begin{tabular}{| c | c | c | c | c | c | c | c |}
\hline
\hline
Compounds & $T_{c}$ & $H_{c1}$ (Oe) & $H_{c2}$ (kOe) & $\xi$ (nm) & $\lambda$ (nm) & $\kappa$ & anisotropy\\
\hline
\hline
FeCr$_{0.02}$Se ($H \parallel (101)$) & 10.5 & 66 & 220 & 3.87 & 334 & 86.3 & Isotropic\\
\hline
$\beta$-FeSe \cite{lei35} & 8.5 & 75(1) & 180 & 4.28 & $\sim$ 309 & 72.3 & Isotropic\\
\hline
FeSe$_{1-x}$Te$_x$ \cite{kim34} & 14 & $<$ 100 & 650 & 2.2 & 560 & 254 & 3.1(2)\\
\hline
FeTe$_{0.8}$S$_{0.2}$ \cite{hu34} & 8.4 & - & 440 & 2.7 & - & - & 1.05(4)\\
\hline
\hline
%\label{tab:table2}
\end{tabular}
\end{center}
\end{table}

\newpage

\begin{table}[!t]
\begin{center}
\caption[Power law decay of activation energy ($U_0 \sim H^{-\alpha}$) for FeCr$_{0.02}$Se single crystals. The values of $\alpha$ for Bi$_{2.2}$Sr$_2$Ca$_{0.8}$Cu$_2$O$_{8+\delta}$, $\beta$-FeSe and SmFeAsO$_{0.85}$ are also provided for a comparison.]{Power law decay of activation energy ($U_0 \sim H^{-\alpha}$) for FeCr$_{0.02}$Se single crystals. The values of $\alpha$ for Bi$_{2.2}$Sr$_2$Ca$_{0.8}$Cu$_2$O$_{8+\delta}$, $\beta$-FeSe and SmFeAsO$_{0.85}$ are also provided for a comparison.}
\begin{tabular}{| c | c | c | c | c |}
\hline
\hline
{Sample} & {Field Orientation} & $H_{\rm cr}$ (kOe) &
\multicolumn{2}{c|} {$\alpha$}\\
\cline{4-5}
{} & {} & {} & $(H < H_{\rm cr})$ & $(H > H_{\rm cr})$\\
\hline
\hline
FeCr$_{0.02}$Se & $\parallel (101)$ & 39.3 & 0.37$\pm$0.03 & 0.88$\pm$0.04\\
\cline{2-5}
{} & $\perp (101)$ &  37.3 & 0.31$\pm$0.02 & 0.85$\pm$0.03\\
\hline
Bi$_{2.2}$Sr$_2$Ca$_{0.8}$Cu$_2$O$_{8+\delta}$ \cite{palstra32, palstra33} & $\parallel a, b$ & 10.0 & 0.48$\pm$0.04 & 0.15$\pm$0.02\\
\cline{2-5}
{} & $\perp a, b$ &  30.0 & 0.16$\pm$0.02 & 0.33$\pm$0.05\\
\hline
$\beta$-FeSe \cite{lei35} & $\parallel (101)$ & 30.0 & 0.25$\pm$0.06 & 0.68$\pm$0.06\\
\cline{2-5}
{} & $\perp (101)$ &  30.0 & 0.26$\pm$0.02 & 0.70$\pm$0.09\\
\hline
SmFeAsO$_{0.85}$ \cite{lee35} & $\parallel c$ & 30.0 & $\approx$ 0.35 & $\approx$ 0.88\\
\cline{2-5}
{} & $\perp (101)$ &  - & - & -\\
\hline
\hline
%\label{tab:table3}
\end{tabular}
\end{center}
\end{table}  

\newpage		

\begin{figure} %fig.1 
\includegraphics[scale=0.4,angle=0]{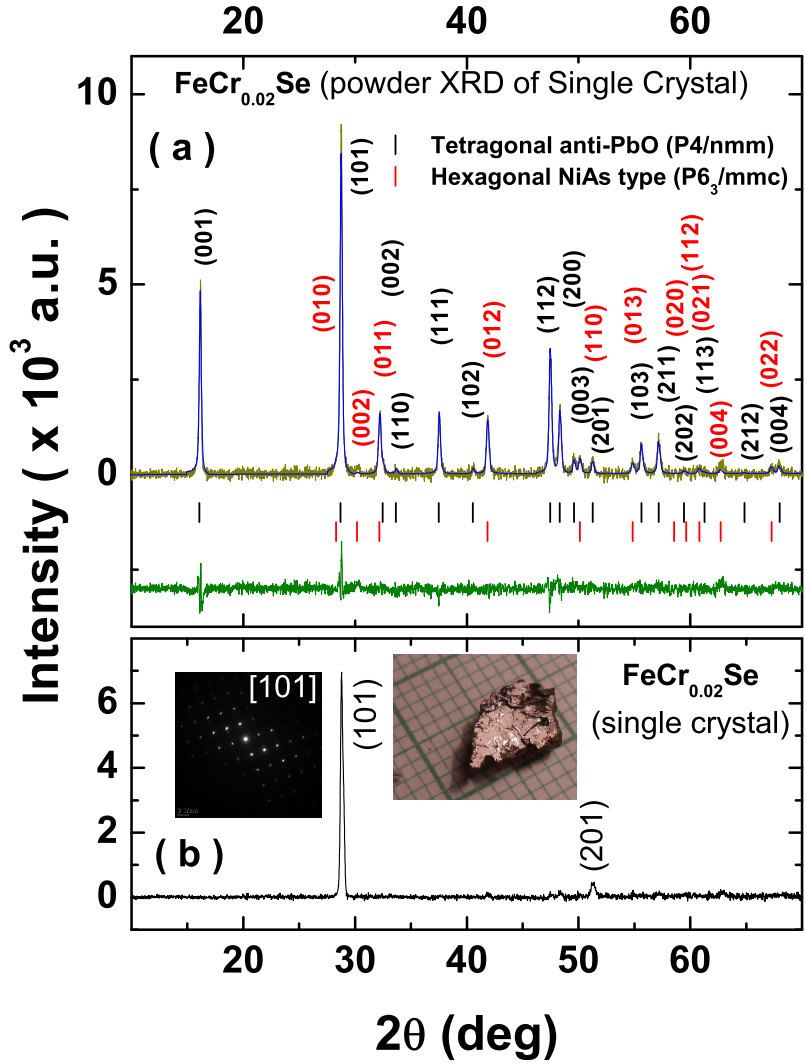}
\caption{(Color online)  (a) X-ray diffraction pattern for powdered single crystals of FeCr$_{0.02}$Se along with the results of the Rietveld analysis. Blue line represents the calculated pattern and the green line represents the difference between observed and calculated patterns. Black and red vertical lines indicate the peak positions of the $\beta$-FeSe and $\alpha$-FeSe phases, respectively. (b) X-ray diffraction pattern for a cleaved piece of single crystal of FeCr$_{0.02}$Se.  ($l$01) lines are identified to show the orientation of the crystal (see text for details). Insets in panel (b) shows a SAED pattern for the crystal as well as an optical micrograph.}
\end{figure}

\begin{figure} %fig.2
\includegraphics[scale=0.4,angle=0]{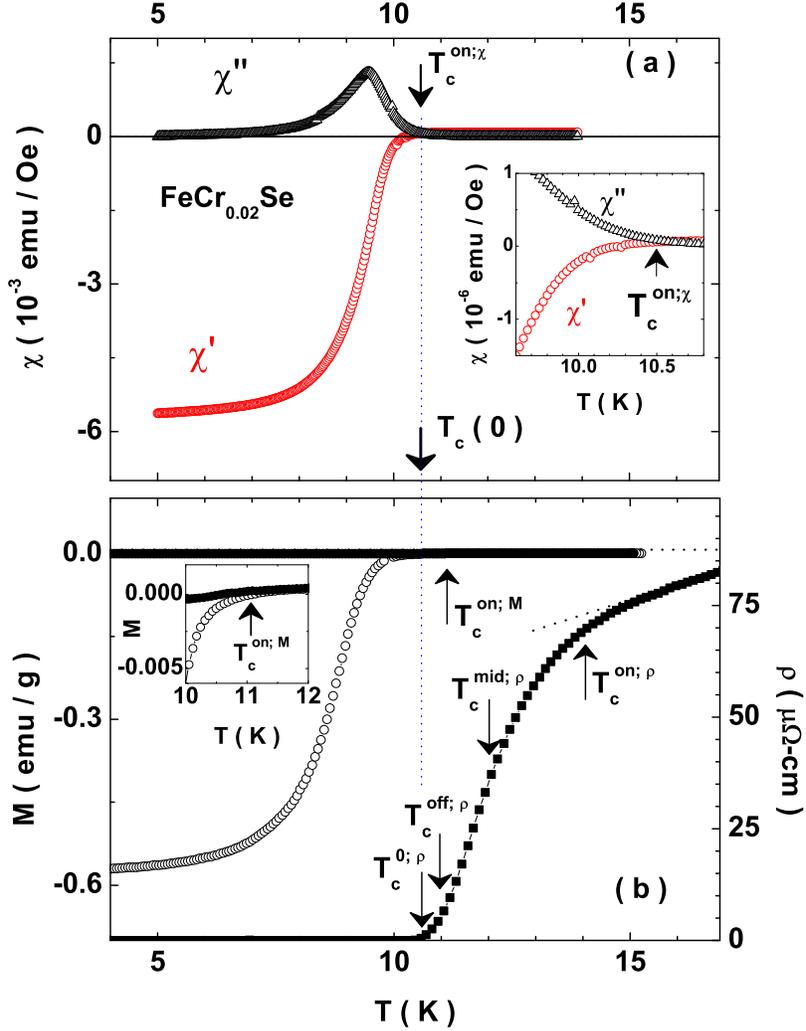}
\caption{(Color online) (a) Temperature variation of the real ($\chi^{\prime}$) and imaginary ($\chi^{\prime\prime}$) parts of ac susceptibility   for a single crystal of FeCr$_{0.02}$Se obtained using $h_{\rm ac} = 3.5$\,Oe at 211\,Hz. Onset temperature $T^{\rm on;\chi}_{c}$ of the superconducting transition is marked. (b) Temperature variation of the dc magnetization $M(T)$ (both ZFC and FC in $H = 10$\,Oe; see text) and resistivity, $\rho(T)$ for a single crystal of FeCr$_{0.02}$Se. The onset temperature based on $M(T)$ measurements, $T^{\rm on;M}_{c}$ is marked by an arrow. Also marked are the onset ($T^{\rm on;\rho}_{c}$), mid ($T^{\rm mid;\rho}_{c}$) and offset ($T^{\rm off;\rho}_{c}$) temperatures obtained at 90\%, 50\% and 10\% of normal state resistivity, respectively. The insets in panel (a) and (b) show the expanded view near T$_{\rm c}$ to identify the temperature more accurately}
\end{figure}

\begin{figure} %fig.3
\includegraphics[scale=0.4,angle=0]{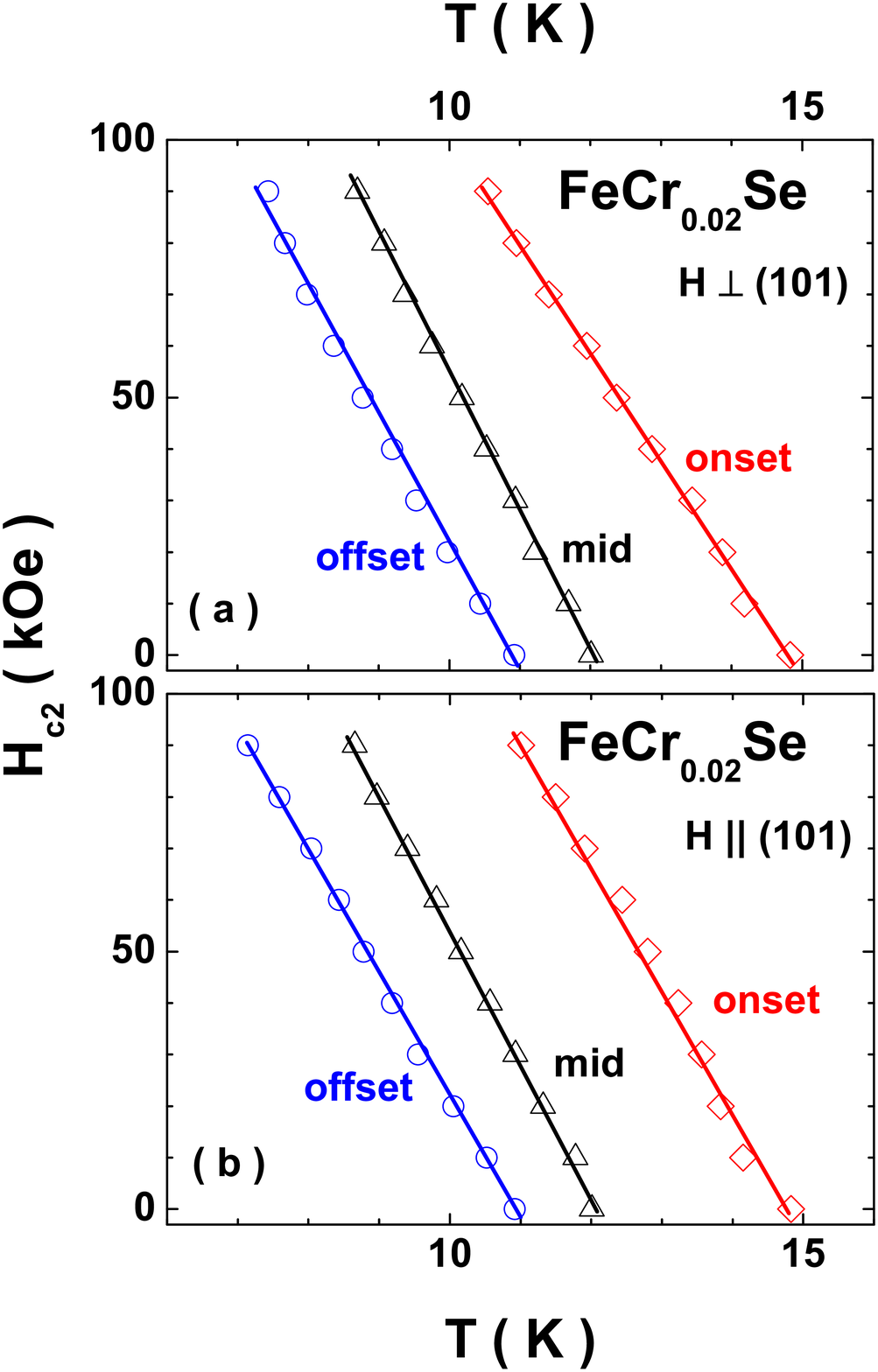}
\caption{(Color online) The upper  critical field, $H_{c2}(T)$ obtained from the field dependent resistivity measurements for $H\parallel (101)$ (panel (a)) and $H\perp (101)$ (panel (b)) using the criteria, the onset (90\%), mid (50\%) and offset (10\%)   of the normal state resistivity. Solid lines are the fit to WHH formalism (see text for details)}
\end{figure}

\begin{figure} %fig.4
\includegraphics[scale=0.6,angle=0]{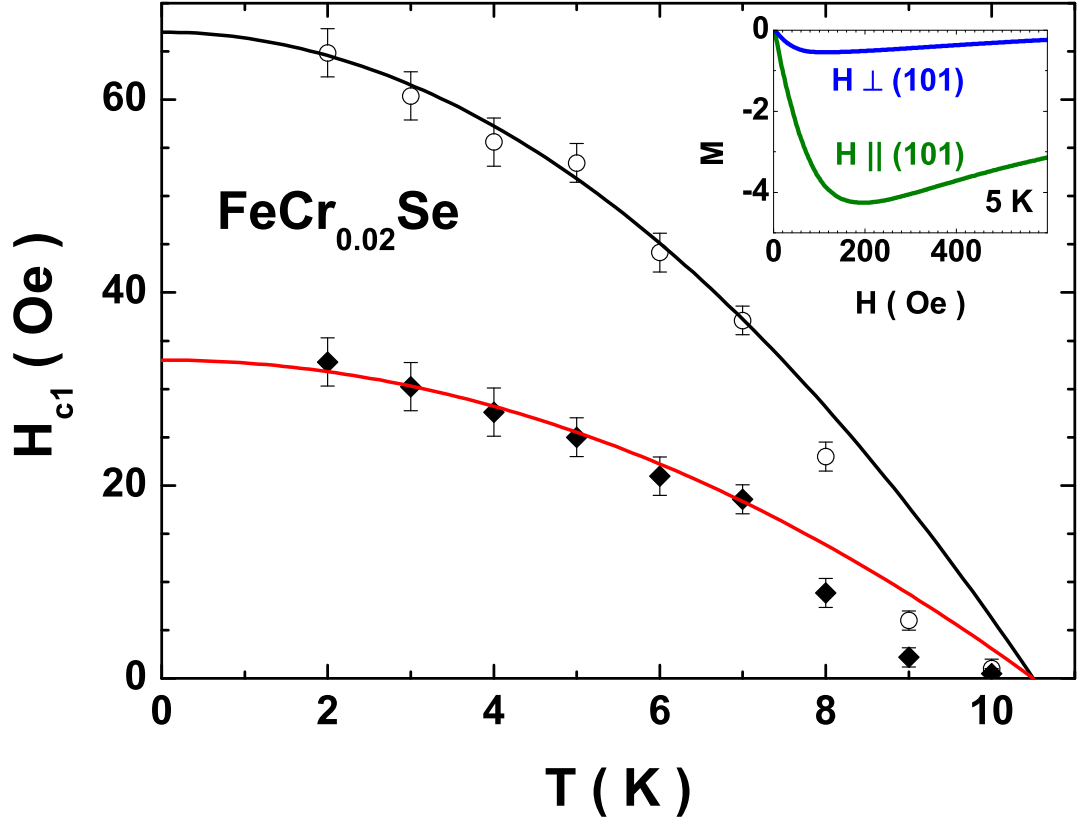}
\caption{(Color online) The lower  critical field, $H_{c1}(T)$ obtained the low field $M$--$H$ measurements for both $H\parallel (101)$ and $H\perp (101)$. Solid lines are obtained from the BCS fit (see text for details). The inset panel shows a portion of the virgin isothermal $M$--$H$ data for both $H\parallel (101)$ and $H\perp (101)$ obtained at 5\,K.}
\end{figure}

\begin{figure} %fig.5
\includegraphics[scale=0.5,angle=0]{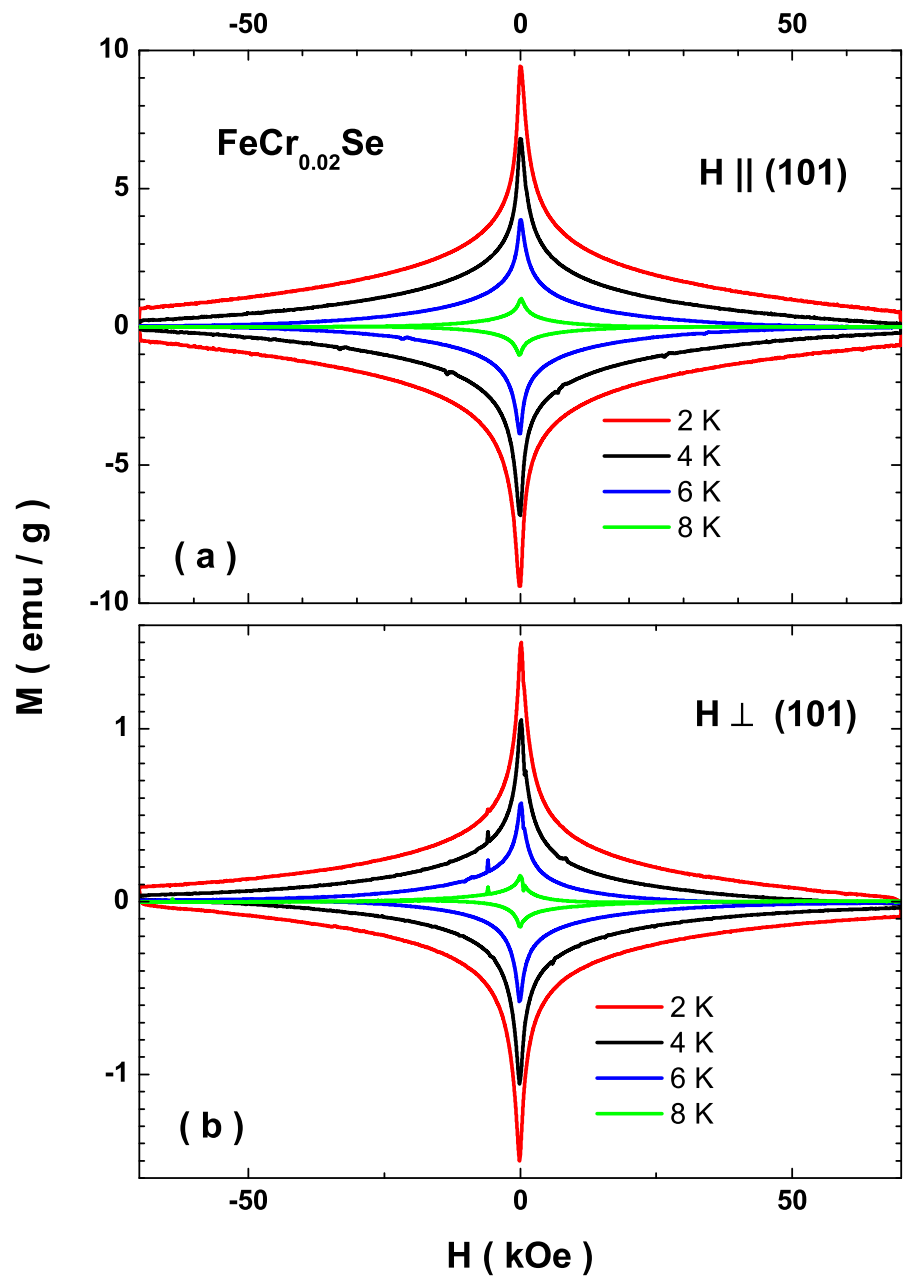}
\caption{(Color online) Isothermal magnetization ($M$--$H$) data obtained at various temperatures for $H\parallel (101)$ (panel (a)) and $H\perp (101)$ (panel (b)) }
\end{figure}

\begin{figure} %fig.6
\includegraphics[scale=0.5,angle=0]{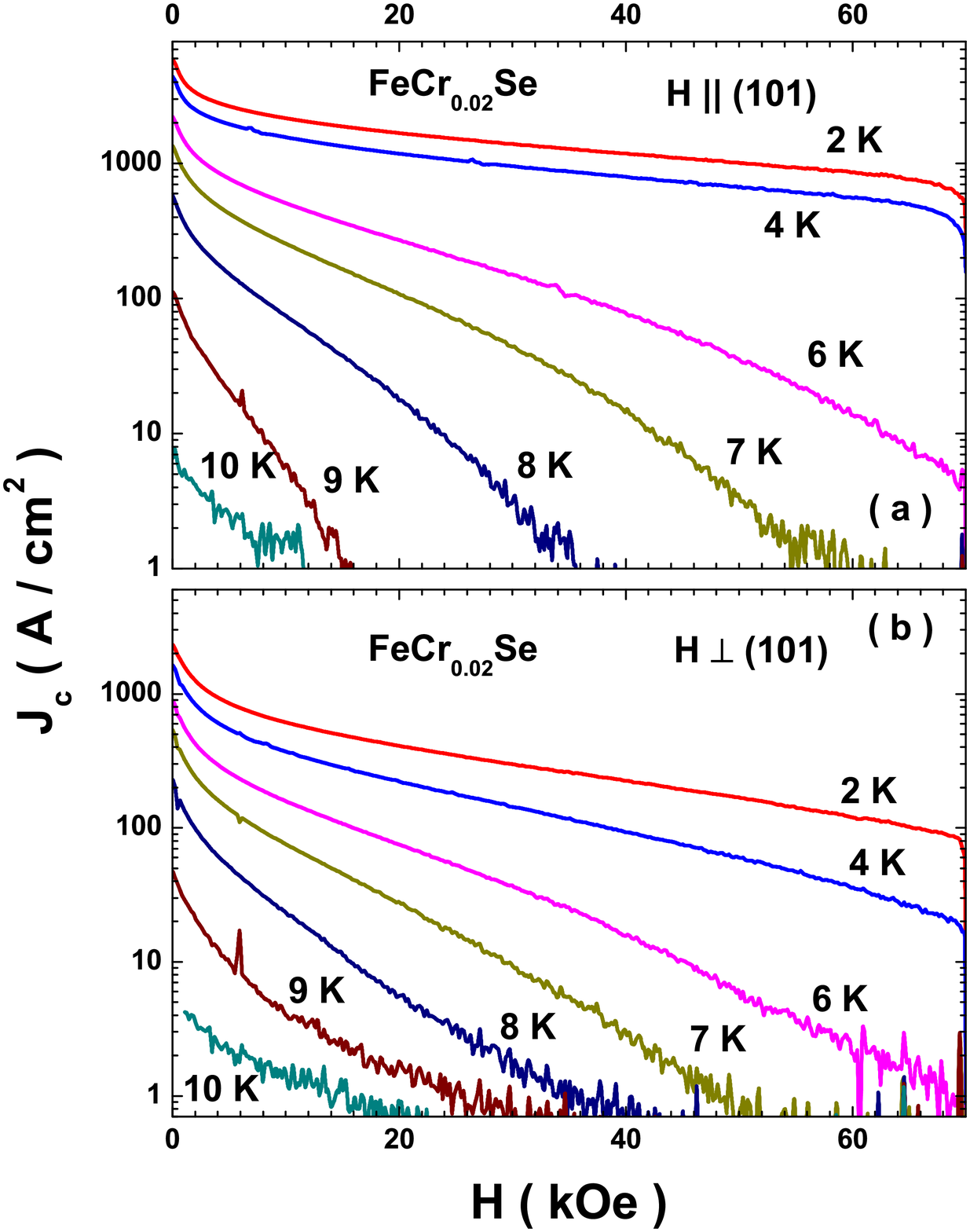}
\caption{(Color online) Critical current density ($J_c(H)$) data obtained at various temperatures for $H\parallel (101)$  (panel (a)) and  $H\perp (101)$ (panel (b)) obtained from the isothermal $M$--$H$ curves of Fig.~5 using the Bean's critical state model (see text for details)}
\end{figure}

\begin{figure} %fig.7
\includegraphics[scale=0.6,angle=0]{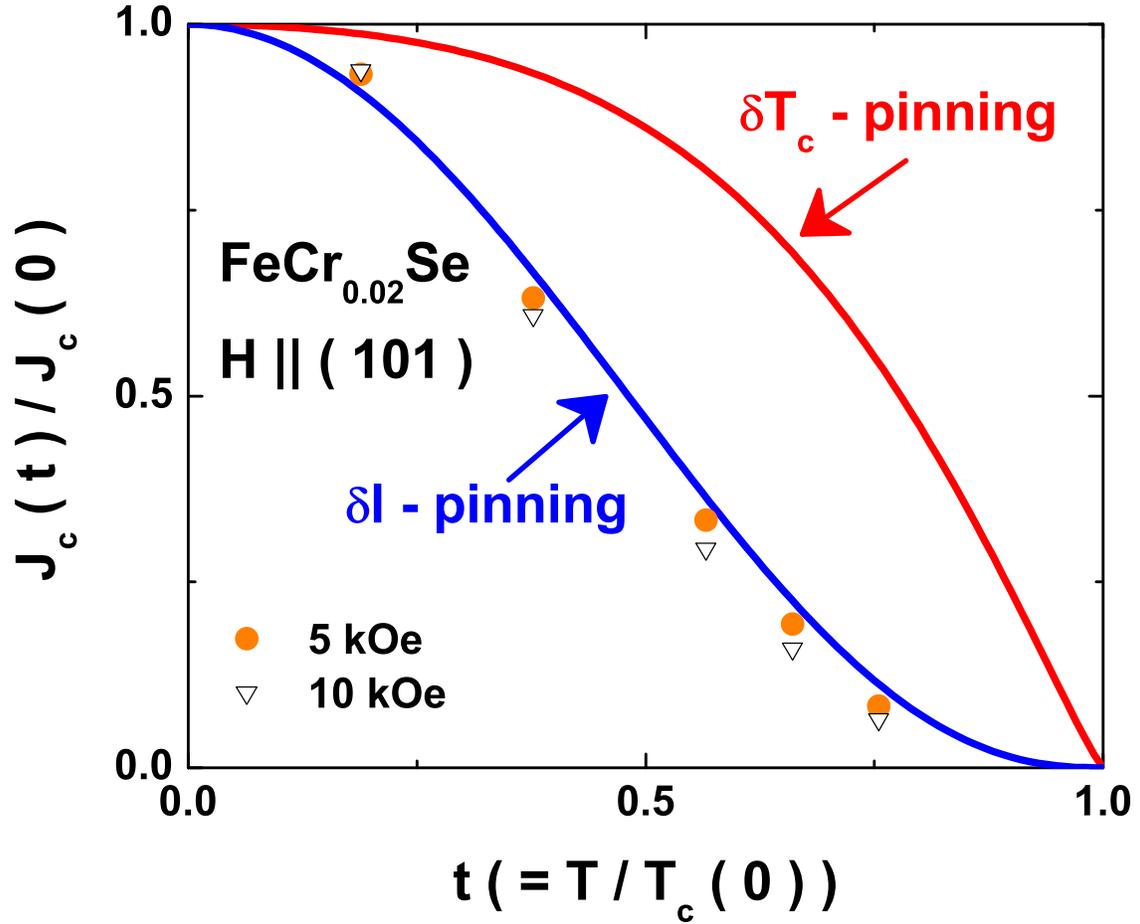}
\caption{(Color online) Theoretically expected variation of the normalized critical current density, ($J_c(t)/J_c(0)$)as a function of reduced temperature $t$ ($=T/T_c$) within the $\delta l$ (blue line) and  $\delta T_c$ (red line) pinning scenario. Also plotted are the experimental values of $J_c(t)/J_c(0)$ obtained at typical magnetic field strengths of 5\,kOe and 10\,kOe for $H\parallel (101)$.}
\end{figure}

\begin{figure} %fig.8
\includegraphics[scale=0.6,angle=0]{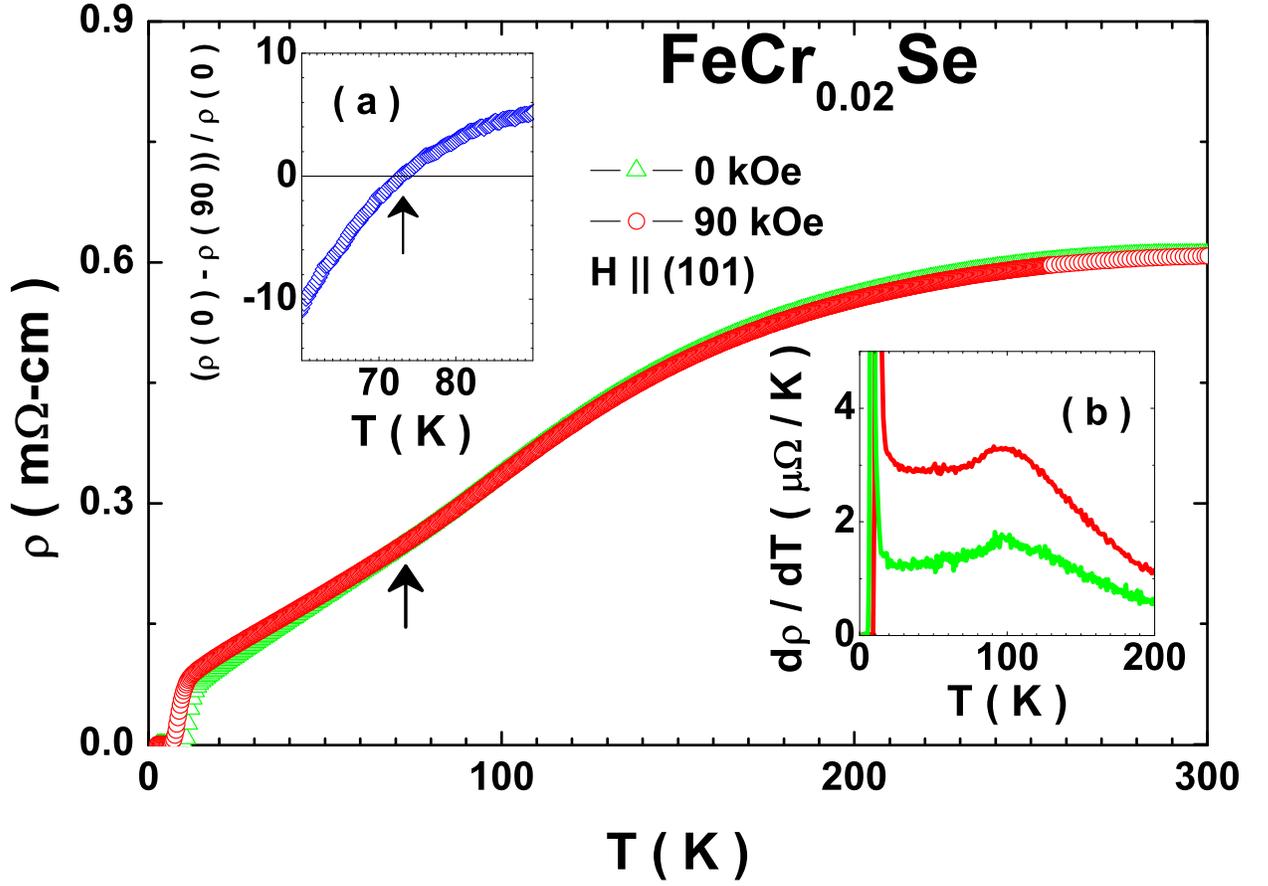}
\caption{(Color online) Variation of resistivity ($\rho$) as a function of temperature $T$ for a suitably cleaved single crystal of FeCr$_{0.02}$Se obtained in  dc magnetic fields of 0\,kOe and 90\,kOe for $H\parallel (101)$. Inset (a): Expanded view of the variation  of magnetoresistance $(\rho (0\,kOe) - \rho (90\,kOe)) / \rho (0\,kOe)$ near 72\,K to highlight a cross over between the two curves (marked by an arrow) Inset (b): Variation of $\frac{d\rho}{dT}$ with temperature highlighting  a hump like behaviour which may be associated with the structural transition in this compound}.  
\end{figure}

\begin{figure} %fig.9 
\includegraphics[scale=0.5,angle=0]{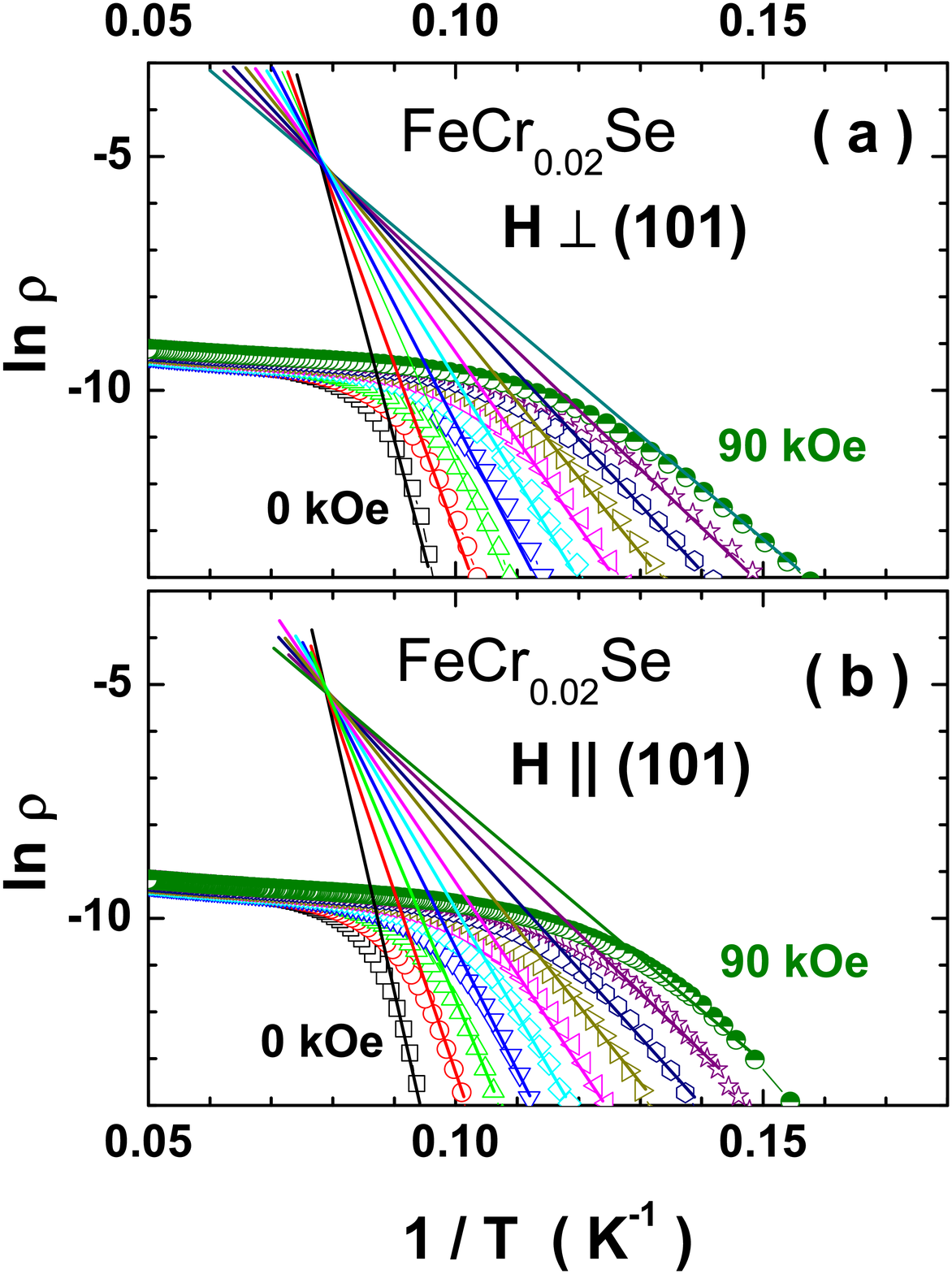}
\caption{(Color online) $\ln\rho$ versus $1/T$ at various  applied field values for (a) $H\parallel (101)$ and (b) $H\perp (101)$. Linear fits to low $T$ data obtained at various  magnetic field values in the range 0--90\,kOe are shown by solid lines in both the panels.}
\end{figure}

\begin{figure} %fig.10
\includegraphics[scale=0.6,angle=0]{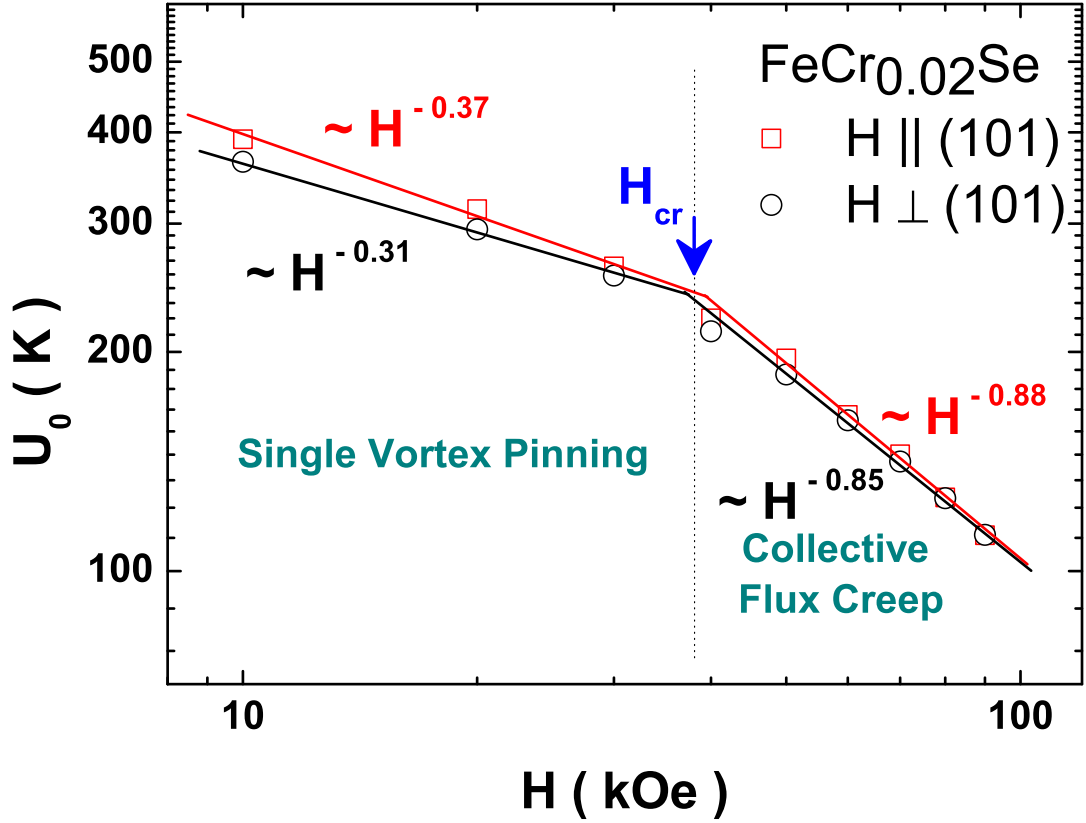}
\caption{(Color online) Activation energy $U_0$ (calculated from $\ln\rho$ versus $1/T$ curves of Fig.~9) versus $H$ for $H\parallel (101)$  and $H\perp (101)$. The power law fits ($H^{-\alpha}$) are shown by lines. Crossover field between single vortex pinning regime and collective flux creep regime, $H_{\rm cr}$ is marked by an arrow.}
\end{figure}

\begin{figure} %fig.11
\includegraphics[scale=0.4,angle=0]{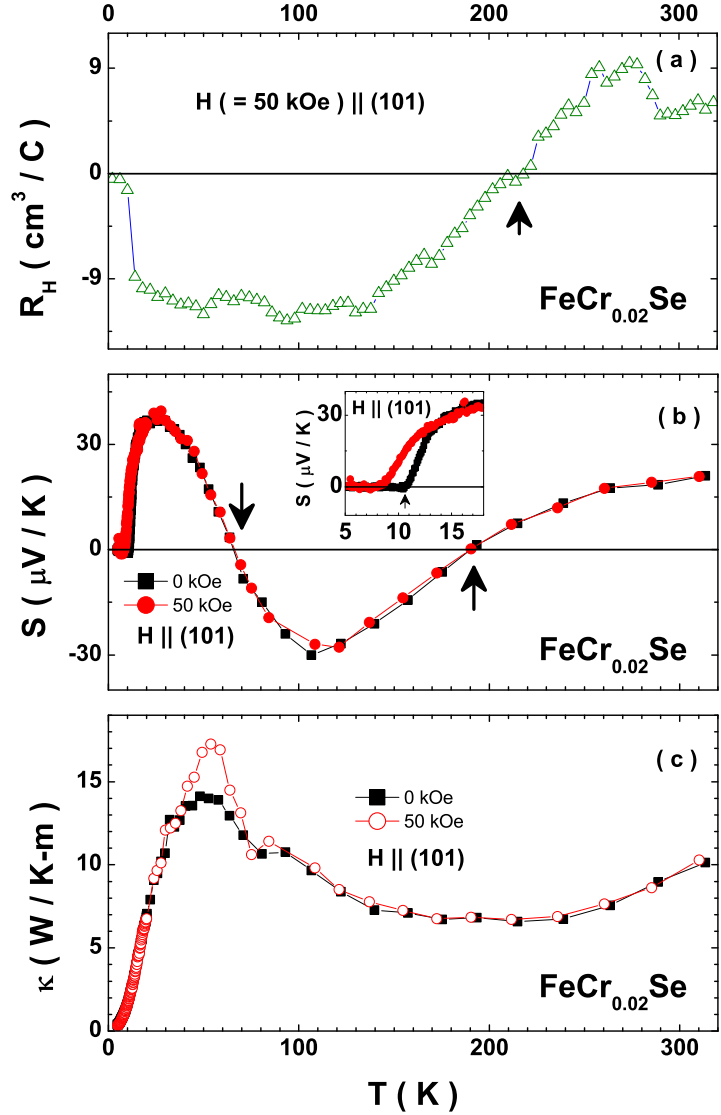}
\caption{(Color online) (a) Variation of Hall coefficient ($R_H$) with temperature of a suitably cleaved single crystal of FeCr$_{0.02}$Se  in a magnetic field of  50\,kOe for $H\parallel (101)$. The temperature at which the sign of the charge carriers change  is shown by an arrow (b) Variation of thermopower ($S$) with temperature  at 0\,kOe and 50\,kOe for $H\parallel (101)$. Temperatures at which $S(T)$ changes sign are shown by arrows. Variation of $S$ with temperature near T$_{\rm c}$ is shown as an inset.  (c) Variation of thermal conductivity ($\kappa$) with temperature  at 0\,kOe and 50\,kOe for $H\parallel (101)$.}
\end{figure}

\begin{figure} %fig.12
\includegraphics[scale=0.6,angle=0]{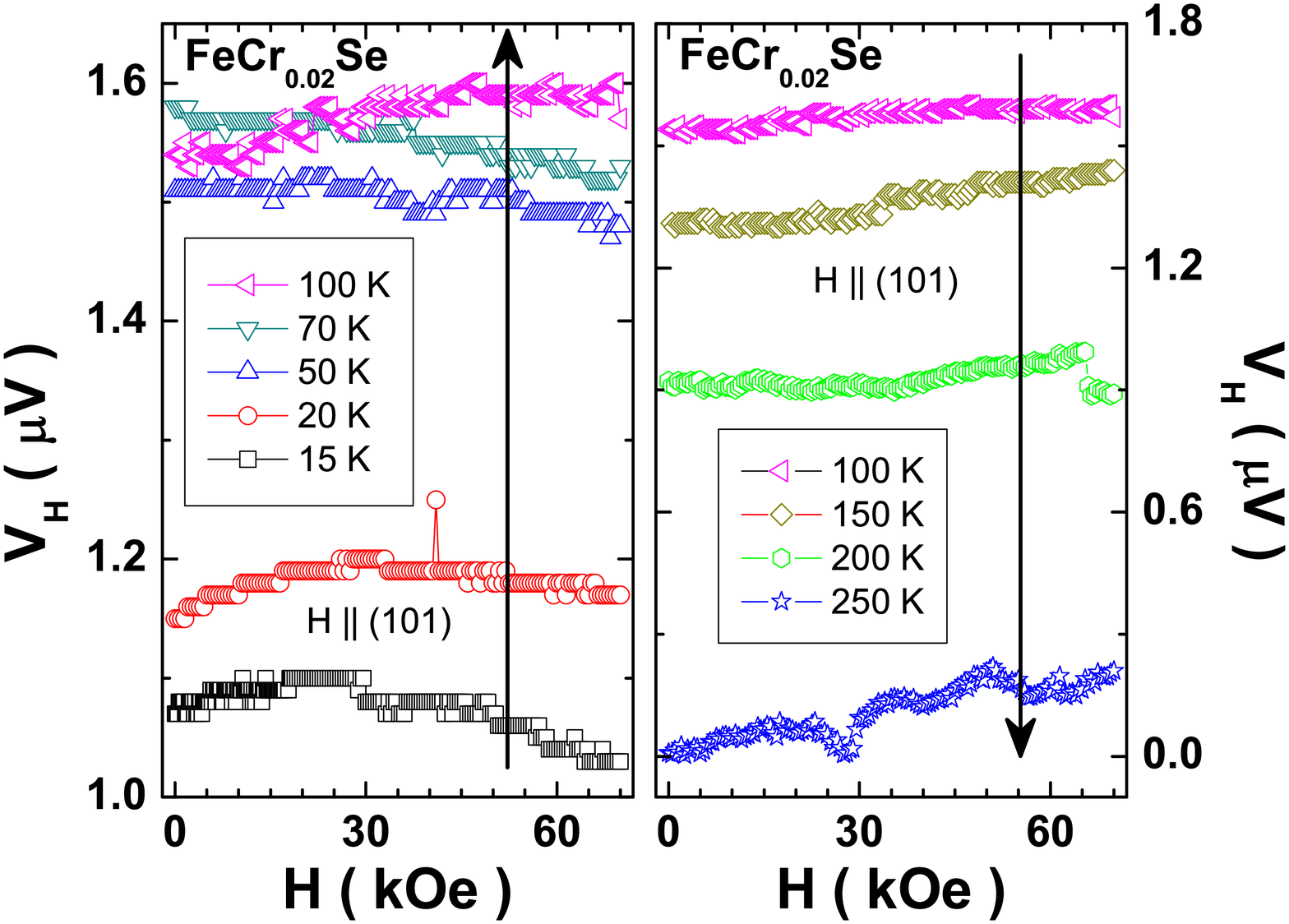}
\caption{(Color online) Variation of Hall coefficient ($R_H$) with magnetic field of a suitably cleaved single crystal of FeCr$_{0.02}$Se  at various temperatures in the normal state for $H\parallel (101)$. The arrows indicate the change in the magnitude of the Hall voltage as the temperature is increased from 15\,K to 250\,K}
\end{figure}
 
\end{document}